\documentstyle[12pt]{article}

\newcommand{\be}{\begin{equation}}
\newcommand{\ee}{\end{equation}}
\newcommand{\bea}{\begin{eqnarray}}
\newcommand{\eea}{\end{eqnarray}}

\newcommand{\p}[1]{(\ref{#1})}

\def\lefthook{{\vrule height5pt width0.4pt depth0pt}}
\def\righthook{{\vrule height5pt width0.4pt depth0pt}}
\def\leftrighthookfill{$\mathsurround=0pt \mathord\lefthook
     \hrulefill\mathord\righthook$}
\def\underhook#1{\vtop{\ialign{##\crcr$\hfil\displaystyle{#1}\hfil$\crcr
      \noalign{\kern-1pt\nointerlineskip\vskip2pt}
      \leftrighthookfill\crcr}}}

\newcommand{\dif}{\partial}

\font\ninerm=cmr9

\begin{document}
\thispagestyle{empty}
\setcounter{page}{0}
\renewcommand{\thefootnote}{\fnsymbol{footnote}}
\fnsymbol{footnote}

\phantom{NOTHING}
\rightline{ITP-SB-96-22\break}
\rightline{INS-Rep.-1140\break}
\rightline{hep-th/9605102\break}
\rightline{May  1996\break}
\vspace{.3in}

\begin{center} \Large \bf An explicit construction of  
Wakimoto  realizations\\

of  current algebras\\

\end{center}

\vspace{.2in}

\begin{center}
Jan de Boer${}^\dagger$ 
and L\'aszl\'o Feh\'er${}^*$\\

\vspace{0.3in}

{\em 
${}^\dagger$Institute for Theoretical Physics\\
State University of New York at Stony Brook\\
Stony Brook, NY 11794-3840, USA\\
e-mail: deboer@insti.physics.sunysb.edu\\

\vspace{0.3in}

${}^*$Institute for Nuclear Studies, University of Tokyo\\
Midori-cho, Tanashi-shi, Tokyo 188, Japan\\
e-mail: laszlo@ins.u-tokyo.ac.jp
}

\end{center}

\vspace{.4in}

{\parindent=25pt
\noindent

\begin{center} {\bf Abstract} \end{center}

\small

It is known from a work  of Feigin and Frenkel that 
a  Wakimoto type, generalized free field 
realization of the  current algebra 
$\widehat{\cal G}_k$ can be associated with 
each parabolic subalgebra ${\cal P}=({\cal G}_0+{\cal G}_+)$ 
of the Lie algebra ${\cal G}$, where in the standard case 
${\cal G}_0$ is  the Cartan  and ${\cal P}$ is the  Borel subalgebra.
In this letter we  obtain  an explicit formula 
for the  Wakimoto realization in the general case. 
Using  Hamiltonian reduction of the WZNW model, 
we  first derive a Poisson bracket realization of the ${\cal G}$-valued 
current in terms of symplectic bosons belonging to ${\cal G}_+$ 
and a current  belonging to ${\cal G}_0$.
We  then quantize the formula  by 
determining the correct  normal ordering. 
We  also show that the affine-Sugawara stress-energy tensor  takes  the
expected quadratic form in the constituents.
}

\normalsize

\newpage

\renewcommand{\thefootnote}{\arabic{footnote}}
\setcounter{footnote}{0}

\section{Introduction}

\def\al{\alpha}
\def\bt{\beta}
\def\d{\delta}
\def\gm{\gamma}
\def\Om{\Omega}
\def\ttt{\tilde{T}}
\def\nonu{\nonumber \\ {} }
\def\p{\phi}
\def\v{\varphi}
\def\la{\langle}
\def\ra{\rangle}
\def\pa{\partial}
\def\s{\vskip 5mm}
\def\i{\int d^2x\,}
\def\hb{\hfill\break}
\def\t{{\rm Tr}\,}
\def\tr{{\rm Tr}\,}
\def\dg{\delta g\,g^{-1}}
\def\gd{g^{-1}\delta g}
\def\fg{{{\d F}\over{\d g}}}
\def\fj{{{\d F}\over{\d J}}}
\def\hg{{{\d H}\over{\d g}}}
\def\hj{{{\d H}\over{\d J}}}
\def\si{\sigma}
\def\bsi{\bar\sigma}
\def\D{{\cal D}}
\def\G{{\cal G}}
\def\M{{\cal M}}

In 1986 Wakimoto \cite{Wa}  found the following  formula 
for the generating currents of $\widehat{sl}(2)_k$:
\be
I_- =-p,
\quad
I_0 =j_0 -2\left(pq\right),
\quad
I_+=-\left(j_0q\right)+k\dif q +\left(p(qq)\right).
\label{1}\ee
Here $j_0=-i\sqrt{2(k+2)}\dif \phi$, and the constituents 
are free fields whose commutation relations are encoded 
by  the singular operator product 
expansions\footnote{We adopt the
conventions of \cite{BBSS} for OPEs and for normal ordering.} 
(OPEs)
\be 
p\underhook{(z) q}(w) = \frac{-1 }{z-w},
\qquad
\dif\phi \underhook{(z) \dif \phi}(w) = \frac{-1}{(z-w)^2}.
\label{2}\ee
The affine-Sugawara  stress-energy tensor is quadratic in the free fields
\be
\frac{1}{2(k+2)}\left({\rm tr\,}I^2\right)=
-\frac{1}{2}
\left(\dif \phi \dif \phi\right) -\frac{i}{\sqrt{2(k+2)}}\dif^2\phi
-\left(p\dif q\right).
\label{3}\ee
This result raised considerable 
interest because of its usefulness in computing correlation
functions of conformal field theories \cite{FF,BF,BMP,ATY}
 and in describing the 
quantum  Hamiltonian
reduction of $\widehat{sl}(2)_k$ to the Virasoro algebra \cite{BO}.

A detailed study of `Wakimoto realizations'
of arbitrary non-twisted affine Lie algebras 
was then undertaken in \cite{FF,BF,BMP,GMOMS,KS,IKa,IKo}
 making use of 
the observation that at the level of zero modes (\ref{1}) becomes
the differential operator realization of $sl(2)$ given by
\be
I_-\rightarrow {\dif \over \dif q},
\qquad
I_0\rightarrow -\mu_0 +2 q {\dif \over \dif q},
\quad
I_+\rightarrow \mu_0 q - q^2 {\dif \over \dif q},
\qquad 
\forall \mu_0\in {\bf C}.
\label{4}\ee
It is well-known (e.g.~\cite{Ko}) 
that any simple Lie algebra, $\G$, admits analogous 
realizations in terms of 
first order  differential operators of the form 
\be
X\rightarrow \hat X=-F^X_\alpha(q) 
{\partial \over \partial q_\alpha}+ H^X_{\mu_0}(q),
\qquad \forall\, X\in \G,
\label{5}\ee
where $F_\alpha^X$, $H_{\mu_0}^X$ are polynomials.  
In the principal case,  $\alpha$ runs 
over the positive roots and $\mu_0$ is an  arbitrary weight.
The main idea  \cite{FF,BMP,IKo}  for generalizing (\ref{1}) 
was  to  regard the differential operator realization of $\G$
as the zero mode part of the sought after realization of $\widehat{\G}_k$,
which should be obtained by ``affinization'' where
one replaces $q_\alpha$ and $p^\alpha$ by conjugate quantum fields and
$\mu_0$ by a current that 
(in the principal case) belongs to the Cartan subalgebra of $\G$.
Of course, one also needs to add derivative terms  and 
find the correct normal ordering, which is rather nontrivial.

Feigin and Frenkel \cite{FF}  demonstrated by indirect, homological 
techniques  that 
the affinization can be performed for any  simple Lie algebra $\G$.
An explicit  formula for the currents 
corresponding to the Chevalley generators of $sl(n)$ was given in \cite{FF} 
too, but  the method does  not lead to  explicit formulas 
for all currents of an  arbitrary $\G$ (or  $\G=sl(n)$).
Some  explicit formulas were later obtained in \cite{IKo}  for any $\G$, 
but only for the Chevalley generators and without complete proofs.
The quadraticity of the affine-Sugawara stress-energy tensor 
in the free fields is not quite transparent in this approach.

In this letter we re-examine the construction of Wakimoto realizations.
As our main result, 
we shall derive  an  explicit formula for the full $\G$-valued 
current in terms of finite group theoretic data.
We also  verify the quadraticity  of  the  affine-Sugawara
    stress-energy tensor.
We shall do this in the  general case
for which the Borel subalgebra 
of $\G$ that features in the principal case is replaced 
by an arbitrary parabolic subalgebra.
The general case was previously  investigated by Feigin and Frenkel 
 (see the second article in \cite{FF}) in their homological approach, 
but we will use a different, direct  method for the construction.

In our method   we first derive a classical, Poisson
bracket (PB) version of the Wakimoto realization of $\widehat{\G}_k$.
This will  naturally result from a Hamiltonian reduction of the WZNW
model based on the Lie group $G$ corresponding to $\G$.
To motivate the construction in a simpler context, 
we shall also consider the finite dimensional 
version of the Hamiltonian reduction, which yields the  PB
analogue  of the differential operator realization of $\G$.
The PB realization of  $\widehat{\G}_k$
is much closer to the desired OPE realization  than 
the differential operator realization of $\G$ is, and it will not
be very difficult to quantize it explicitly by normal ordering.
The formula  that we shall obtain 
(given in (\ref{qwak}) together with  (\ref{omega}))
 should be useful in future applications of
the Wakimoto realization. 

In this letter we explain  the construction outlined 
in the preceding paragraph and announce the main result;
some proofs will be deferred to a subsequent paper \cite{dBF}.

In the rest of this section we collect some notions of Lie theory  
(see e.g.~\cite{GOV}) that will be used. 
Let the step operators
$E_{\pm \alpha_l}$ and the Cartan elements $H_{\alpha_l}$,
associated  with the simple roots 
$\alpha_l$ for $l=1,\ldots, r={\rm rank}(\G)$, 
be the  generators of the complex simple Lie algebra $\G$.
For any $l=1,\ldots, r$,  choose an integer
$n_l\in \{ 0,1\}$ and determine the unique Cartan element $H$ for which 
$[H, E_{\pm \alpha_l}]=\pm n_l  E_{\pm \alpha_l}$
(choices related by symmetries of the Dynkin diagram are equivalent).
The eigenspaces of $H$ in the adjoint representation 
define a ${\bf Z}$-gradation of $\G$, 
\be
\G =\oplus_m\,  \G_m
\qquad
[\G_m, \G_n] \subset \G_{m+n}
\qquad\hbox{with}\quad
\G_m\equiv\{\, X\in \G\,\vert\, [H, X] = mX\,\}.
\label{grading}\ee 
Denoting  the subspaces of positive/negative grades  by $\G_\pm$,
we  obtain the decomposition 
\be 
\G=\G_- + \G_0 + \G_+.
\label{triang}\ee 
We also consider a connected complex Lie group $G$ 
whose Lie algebra is $\G$ and its connected subgroups
 $G_{0,\pm}$ corresponding to the subalgebras $\G_{0,\pm}$,
respectively.
We then have the dense  open submanifold $\check G\subset G$ 
of `Gauss decomposable' elements,
\be
\check G\equiv\{\, g=g_+g_0 g_-\,\vert \, g_{0,\pm}\in G_{0,\pm}\,\},
\label{gauss}\ee
which equals to  $G_+\times G_0 \times G_-$ as a  manifold
since the decomposition of any $g\in \check G$ is unique.
The {\em parabolic}  subalgebra ${\cal P}\subset \G$ associated with 
the fixed set of integers $n_l$, 
and the corresponding parabolic subgroup $P\subset G$ are given by
\be
{\cal P}\equiv(\G_0+\G_+),
\qquad \quad
P=G_0G_+,
\ee
and have a semidirect product structure since $[\G_0,\G_+]=\G_+$.
For $X, Y\in \G$, we shall denote an  invariant scalar product 
$\langle X,Y\rangle$ simply as $\tr(XY)$ 
as if a matrix representation of $\G$ was chosen. 
Similarly,  we denote say ${\rm Ad\,} g (X)$ as
$gXg^{-1}$  for any $g\in G$, $X\in \G$.
This notation is used purely for convenience, 
a choice of representation is never needed below.
In the principal case, for which $n_l=1$ $\forall l$, 
 $\G_\pm$ are the subalgebras generated by
the positive/negative roots, $\G_0$ is the Cartan subalgebra,  and 
the parabolic subalgebra is the Borel subalgebra.
In the general case, the Lie algebra ${{\cal G}}_0$ 
can  be decomposed  into an abelian 
factor, say  $\G_0^0$,
and simple factors, say $\G_0^i$ for $i>0$,  that are orthogonal 
with respect to $\tr$,
\be {\cal G}_0 = \oplus_{i\geq 0}\, {\cal G}_0^i,  
\label{reductive}\ee
and $\cal P$ contains the Borel subalgebra.
If $\psi$ and $\psi_i$ denote the  highest roots 
of $\G$ and  of $\G_0^i$ for $i>0$, then the dual Coxeter numbers
of $\G$ and of $\G_0^i$ are respectively given by 
\be
h^*= {c_2(\G)\over \vert\psi\vert^2}\quad\hbox{and}\quad   
h^*_i = {c_2(\G_0^i)\over \vert\psi_i\vert^2} \quad\hbox{for}\quad  i>0.
\label{Cox}\ee
The quadratic Casimir $c_2(\G)$ of $\G$ is defined by
$\eta^{ab} [T_a,[T_b,Y]] = c_2(\G) Y $
where $Y\in\G$, $\eta^{ab}$ is the inverse of $\eta_{ab}=\tr(T_aT_b)$
for a basis $T_a$ of $\G$; 
and $c_2(\G_0^i)$  is defined  analogously.

\section{Hamiltonian reduction in the finite group case}
\setcounter{equation}{0}

We next derive a PB realization of the Lie algebra $\G$,
which is the zero mode part of the classical Wakimoto realization of 
$\widehat{\G}_k$.
The formula  given by eq.~(\ref{finwak}) below
will follow  from a Hamiltonian symmetry 
reduction\footnote{As a general reference, see e.g.~\cite{AM}.}  
of the holomorphic cotangent bundle $T^*G$ of $G$.

Using right translation by the elements $g\in G$, we have the  
identification 
\be
T^*G= G\times \G=\{\, (g,J)\, \vert \, g\in G,\, J\in \G\,\}.
\ee
The canonical symplectic form $\omega$ of $T^*G$ is then given by
\be
\omega = d\, {\tr}\left(J dg g^{-1}\right)
\ee
where $d$ denotes the  exterior derivative.
There are two commuting actions of $G$ on $T^*G$, which are induced by 
left and right translations on $G$ by any group element.
The infinitesimal generators (the momentum map) that generate 
the left translations through PBs are just the components 
of $J$, while the infinitesimal generators of the right translations are
the components of
\be
I= -g^{-1} J g.
\ee
Both the components of $J$ and $I$ satisfy the Lie algebra
$\G$ under the PB defined by  $\omega$. 
We now  consider a symmetry reduction using the action of the parabolic
subgroup $P\subset G$ given by the left translations
\be
L_p: (g,J)\mapsto (pg, pJp^{-1})
\qquad p\in P.
\label{Pact}\ee 
Decomposing $J$ as $J=(J_-+J_0+J_+)$ according to (\ref{triang}),
the momentum map corresponding to (\ref{Pact}) is the projection
$(J_-+J_0)$, and we define the reduction by imposing the constraints
\be
J_-=0 \quad\hbox{and}\quad J_0=\mu_0
\label{constr}\ee
with some arbitrarily chosen $\mu_0\in \G_0$.
The first class part of the constraints generates  the subgroup 
$P(\mu_0)\subset P$, which is easily seen to have the structure 
\be
P(\mu_0)=G_0(\mu_0) G_+
\qquad\hbox{with}\qquad
G_0(\mu_0)\equiv\{\, g_0\in G_0\,\vert\, 
g_0 \mu_0 g_0^{-1}=\mu_0\,\}.
\ee
We are only interested in the dense open submanifold 
of the reduced phase space which results by restricting the group 
element $g\in G$ to belong to $\check G\subset G$ in (\ref{gauss}).
We find that a complete set of gauge invariant (i.e.~$P(\mu_0)$ invariant)
functions on the constrained manifold
\be
\left(T^* \check G\right)_{\rm const}=\{\, (g_+ g_0 g_-, J_+ +\mu_0)\,
\vert\, g_{0,\pm}\in G_{0,\pm},\,\,\, J_+\in \G_+\,\}
\ee
is given by the functions 
\be
g_-,\qquad
j\equiv g_0^{-1}\left(g_+^{-1}(J_+ +\mu_0)g_+ -\mu_0\right) g_0,
\qquad
j_0\equiv - g_0^{-1} \mu_0 g_0.
\ee
Furthermore, the mapping 
\be
(g_-, j, j_0): \left(T^* \check G\right)_{\rm const}\rightarrow 
G_-\times \G_+ \times {\cal O}(-\mu_0)
\ee
gives rise to a corresponding  holomorphic diffeomorphism 
\be  
(g_-, j, j_0): \left(T^* \check G\right)_{\rm red}\equiv  
P(\mu_0)\backslash \left(T^* \check G\right)_{\rm const}
\rightarrow  T^* G_- \times {\cal O}(-\mu_0),
\ee
where the identification $T^* G_-=G_- \times \G_+$ is made 
by right translations and 
${\cal O}(-\mu_0)\equiv G_0(\mu_0)\backslash G_0$ is the
 co-adjoint orbit of $G_0$ through $-\mu_0$.
The canonical symplectic form on $T^* G_- \times {\cal O}(-\mu_0)$ 
coincides with the induced symplectic form  
on the reduced phase space 
since
\be
\omega = d\, {\tr}\left(J dg g^{-1}\right)
= d\, {\tr}\left(j dg_- g_-^{-1}\right) 
+d\, {\tr}\left( \mu_0 dg_0 g_0^{-1}\right)
\quad\hbox{on}\quad 
\left(T^* \check G\right)_{\rm const}.
\label{redsymp}
\ee
Indeed, the two terms  on the right hand side are recognized as  
the symplectic form of $T^*G_-$ and the symplectic form of 
${\cal O}(-\mu_0)$. 

The  point of our construction is that, as it is invariant 
under the symmetry (\ref{Pact}), $I$ survives the reduction.
We have
\be
I = -(g_+ g_0 g_-)^{-1} (J_+ +\mu_0) (g_+ g_0 g_-)=
g_-^{-1} ( -j + j_0) g_-
\quad\hbox{on}\quad 
\left(T^* \check G\right)_{\rm const}.
\ee
Hence the result is the formula
\be
I(g_-, j, j_0)= g_-^{-1} ( -j + j_0) g_-
\label{finwak}\ee
yielding  a realization of the PB algebra of $\G$  
\be
\{ \tr(X I)\,,\, \tr(Y I)\}=\tr( [X,Y]I)
\qquad
X, Y\in \G 
\ee
in terms of the PB algebra of the constituents $g_-, j, j_0$.
Introducing some global, holomorphic coordinates 
$q_\alpha$ on $G_-$ and corresponding canonical coordinates
$(q_\alpha, p^\alpha)$ on $T^*G_-$, 
${\tr}\left( jdg_- g_-^{-1}\right) = p^\alpha dq_\alpha$,
the PB algebra of the constituents becomes
 \be
\{ q_\alpha, p^\beta\} = \delta_\alpha^\beta,
\qquad
\{ \tr(X_0 j_0)\,,\, \tr(Y_0 j_0)\}=\tr( [X_0,Y_0]j_0)
\qquad
X_0, Y_0\in \G_0.
\label{const}\ee
If we further let $V_\alpha\in \G_-$ and $V^\alpha\in \G_+$ 
denote dual bases, $\tr(V_\alpha V^\beta) =\delta_\alpha^\beta$,
then we have\footnote{Summation on 
coinciding indices is understood throughout the text.}
\be
j(q,p) = N^{-1}_{\al\beta}(q) V^{\al}p^{\beta}
\quad\hbox{with}\quad
N^{\alpha\beta}(q)\equiv
\tr\left(V^\beta {\pa g_- \over  \partial q_\alpha}g_-^{-1}\right),
\label{jeq}\ee
since  ${\tr}( V_\alpha j)$  generates  the action of $V_\alpha\in \G_-$
on $T^* G_-$ that comes from left translations,
as is clear from (\ref{redsymp}).
In summary, 
substituting $g_-(q)$ and $j(q,p)$ in  (\ref{finwak}),
we obtain a realization of the PBs of $\G$ in terms of the simpler 
PBs in (\ref{const}). 
Note that the  differential operator realization
of $\G$ on $P\backslash \check G$ mentioned in (\ref{5})
can be recovered by quantizing this PB realization.
In the case when $\mu_0$ is a character of $\G_0$,
(\ref{finwak}) is in fact equivalent to the formula of 
the differential operator realization  given by Theorems 2.1 and 2.2 in 
\cite{Ko}.

\section{Classical Wakimoto realizations of $\widehat{\G}_k$}
\setcounter{equation}{0}

We here present the affine Lie algebraic analogue of the PB realization
of the Lie algebra $\G$ derived in the preceding section.
The PBs of $\widehat{\G}_k$  are given by
\be
\{\,\t (X I)(\si) \,, \, \t (Y I)(\bsi) \,\}
     = \t ([X,Y]   I)(\si) \,\delta
                -K\, \t (X\, Y) \, \d' 
\qquad   X,Y\in \G,
\label{KMPB}\ee
where $I\in C^\infty(S^1, \G)$ is a $\G$-valued  current
and $\delta\equiv\d(\si-\bsi)$ with  $\sigma\in [0,2\pi]$ being 
a coordinate on $S^1$. 
In fact, 
the proper  analogue of formula (\ref{finwak}) turns out to be
\be
I(q,p,j_0)=g_-^{-1}\left(-j + j_0\right)g_- 
+K g_-^{-1} g_-'
\label{affwak}\ee
where $j$ is still given by  (\ref{jeq}) but 
the constituents are  now promoted to be fields on $S^1$.
More precisely, the 
$\G_0$-valued current $j_0$  and the fields $q_\alpha$, $p^\beta$ 
are subject to  the PBs 
\bea
\{\,\t (X_0 j_0)(\si) \,, \, \t (Y_0 j_0)(\bsi) \,\}
     &=& \t ([X_0,Y_0]  j_0)(\si) \,\delta
                -K\, \t (X_0\, Y_0) \, \d' 
\quad   X_0,Y_0\in \G_0,
\nonumber\\ 
\{ q_\alpha(\sigma), p^\beta(\bar \sigma)\} 
&=& \delta^\beta_\alpha\delta(\sigma-\bar \sigma).
\label{pqPB}\eea
Using that $g_-$ depends on the $q_\alpha$ 
by means of the chosen parametrization of $G_-$,
one  can write the PBs in (\ref{pqPB}) equivalently as
\bea
\{\,\t (V_\alpha\, j)(\si) \,, \, \t (V_\beta\, j)(\bsi) \,\}
     &=& \t ([V_\alpha,V_\beta]\, j)(\si) \,\d,
\nonumber \\
\{\,\t (V_\alpha\, j)(\si) \,, \, g_-(\bsi) \,\}
     &=& - V_\alpha\, g_-(\si) \,\d.
\eea
The  result  that the  PBs of $\widehat{\G}_k$ in (\ref{KMPB})
follow from the  PBs of the constituents via the
 formula in (\ref{affwak})
may be  verified   by a straightforward calculation.
It can also be  checked 
that the affine-Sugawara stress-energy 
tensor has 
the following 
quadratic  structure:
\be
{1\over 2K} \t (  I^2) = {1\over 2K}\t (j_0^2) - 
\t (j g_-' g^{-1}_-)={1\over 2K}\t (j_0^2) 
-  p^\alpha q_\alpha'.
\label{clsug}\ee

It is natural to derive formula (\ref{affwak})  by Hamiltonian symmetry
reduction of the WZNW model \cite{Wi}  based on $G$,
generalizing the construction presented in the finite group case.  
We take  the  phase space of the WZNW model  to be  
the holomorphic cotangent bundle of the loop group $\widetilde G$ of $G$, 
which we realize  as  
\be
T^* {\widetilde G}_K= \{\, (g,J)\,\vert\,
g\in C^\infty(S^1,G),\,\,\, J\in C^\infty(S^1,\G)\,\}.
\ee
This phase space has the symplectic form 
\be
\omega_K= \int_{S^1} \delta\, {\rm Tr}\left( J \delta g g^{-1}\right) 
+ K\int_{S^1}
 {\rm Tr}\left(\delta g g^{-1}\right) \left(\delta g g^{-1}\right)' 
\ee
where  $\delta$ denotes 
the functional exterior derivative (see e.g.~\cite{HK}). 
Note that the `left current' $J$ satisfies the analogue of (\ref{KMPB})
with opposite central term. 
The constraints we need to impose are  formally identical 
to those in (\ref{constr}). 
Assuming that $g(\sigma)\in \check G$, 
on the constrained manifold we find the gauge invariant 
`right current' 
\be
I=-g^{-1} J g + Kg^{-1} g' = 
-\left(g_+g_0g_-\right)^{-1}\left(J_+ +\mu_0\right)
\left(g_+g_0g_-\right) + 
K \left(g_+g_0g_-\right)^{-1} \left(g_+g_0g_-\right)',
\ee
which is just  $I$ in  (\ref{affwak}) in terms of the gauge invariant 
objects
\be
j\equiv g_0^{-1} \left( g_+^{-1}\left(J_+ +\mu_0\right)g_+-\mu_0 
-Kg_+^{-1} g_+' \right)g_0,
\qquad
j_0\equiv - g_0^{-1} \mu_0 g_0 +K g_0^{-1}g_0'.
\ee
The PBs of $I$ and those  of $g_-$, $j$, $j_0$ 
follow from this Hamiltonian  reduction (see also \cite{dBF}).

\section{Quantization of the classical Wakimoto realization}
\setcounter{equation}{0}

Our goal now is to derive a quantum counterpart of the
classical Wakimoto realization  (\ref{affwak}). As this classical
realization was derived by means of a Hamiltonian reduction, there seem
a priori to be two ways to quantize it.
The first possibility would require us to write down a quantization of the
phase space of the WZNW model and
subsequently to implement a quantum Hamiltonian
reduction.
Although it might be very interesting to pursue this line of thought
further, it seems to be rather cumbersome, and in addition in the case
at hand it turns out to be relatively easy to directly quantize the
classical Wakimoto realization. Therefore we will restrict ourselves to
the latter method. 

Since the classical Wakimoto realization expresses the currents of $\G$
in terms
of currents in ${\cal G}_0$ and a set of coordinates
and momenta that constitute 
the cotangent bundle $T^*{\widetilde G}_-$
of $\widetilde G_-=C^\infty(S^1, G_-)$, our philosophy will
be to first quantize these objects by postulating 
OPEs  for their generators, and subsequently to write down
a normal ordered version of (\ref{affwak}) in terms of these generators.
The hard work lies  in verifying that the currents defined in this way
indeed satisfy the OPEs of the  affine Lie algebra based on ${\cal G}$.
This requirement will in addition  fix the 
ambiguities that one has to deal with in normal ordering
(\ref{affwak}).

Fixing a basis $T_a$ of $\G$, 
 the OPEs corresponding 
to (\ref{KMPB}) should  read as 
\be
\tr (T_a \underhook{I)(z) \tr(T_b} I)(w)  =
\frac{K \tr( T_a T_b)}{(z-w)^2} +
\frac{\tr( [T_a,T_b] I)(w)}{z-w}.
\label{requi}
\ee
Replacing (\ref{KMPB}) with (\ref{requi})
amounts to replacing the PBs of the Fourier modes of the current   with  
corresponding commutators, as is well-known.
Naturally, 
the OPEs of the constituent coordinate and momentum fields are  
declared to be
\be p^{\al}\underhook{(z) q_{\bt}}(w) = \frac{-\delta^{\al}_{\bt} }{z-w}.
\label{pqOPE}
\ee
Decomposing  the $\G_0$-valued current $j_0$ as $j_0=\sum_{i\geq 0} j_0^i$
according to (\ref{reductive}),
we postulate the OPEs  of the current $j_0^i$ in ${\cal G}_0^i$  as   
\be
\tr (\pi_0^i(T_a) \underhook{ j_0^i)(z) \tr(\pi_0^i(T_b) } j_0^i)(w)  =
\frac{K_0^i \tr( \pi_0^i(T_a) \pi_0^i(T_b))}{(z-w)^2} +
\frac{\tr( [\pi_0^i(T_a),\pi_0^i(T_b)] j_0^i)(w)}{z-w}
\label{j0i}\ee
where $\pi_0^i:{\cal G} \rightarrow {\cal G}_0^i$ is the orthogonal 
projection onto ${\cal G}_0^i$.
All other OPEs of the constituents are regular. 
Note that we have taken the central
extension $K_0^i$ of $j_0^i$ to be a free parameter,
to be determined from requiring (\ref{requi}), 
and that the properly normalized {\em level} parameters of $I$ and  $j_0^i$ 
(which are integers in a  unitary highest weight representation)
are respectively given by  
\be
k\equiv{2K\over \vert \psi\vert^2}
\qquad\hbox{and}\quad
 k_0^i\equiv{2 K_0^i\over \vert \psi_i\vert^2}\quad\hbox{for}\quad i>0.
\label{level}\ee

Notice now that the  classical Wakimoto current in (\ref{affwak}) 
is linear in the $p^\al$ and in $j_0$, but  could contain arbitrary functions 
of the $q_\alpha$ if the coordinates  were not chosen with care. 
However, we here only  wish to deal with objects that are polynomial in the 
basic quantum fields, since those are easily defined in chiral conformal 
field theory 
(which is the same as the theory of vertex algebras \cite{VA})  by 
normal ordering.
Below we will define a class of coordinates 
on $G_-$, the so-called `upper triangular coordinates', in which
the quantum Wakimoto current will be polynomial.
The computations will also simplify considerably in these coordinates.

The  only ordering problem in (\ref{affwak}) arises from 
the term $-g_-^{-1} j g_- = -g_-^{-1} N^{-1}_{\alpha\beta}(q) p^{\beta}
V^{\alpha} g_-$,  for which we have
to choose where to put the momenta $p^{\beta}$.
We  now choose
to put them on the left, and  replace any  classical
object $pf(q)$ by the normal ordered object $(p(f(q)))$. 
{}From the 
OPE\footnote{\ninerm The notations 
${\partial^\alpha f}(q)\equiv
{\partial f(q)\over \partial q_\alpha}$ and  $(\partial F)(z)\equiv
{\partial F(z) \over \partial z}$ are used 
for functions $f$ of $q$ and $F$ of $z$ from now on.}
\be 
p^{\alpha}\underhook{(z) f(q(}w)) = \frac{-\partial^{\alpha} f}{z-w}
\ee
we see 
that $\left([p^{\alpha},f(q)]\right)=-\partial \partial^{\alpha} f(q)
=-\partial^{\beta} \partial^{\alpha} f(q) \dif q_{\beta}$. Hence, 
the difference between two normal orderings of the classical object
$pf(q)$ will always be of the form $\Omega^{\beta}(q) \partial q_{\beta}$
for some function $\Omega^{\beta}$, and we should allow for an additional
term of this type in  quantizing (\ref{affwak}). Altogether this leads
to the following proposal for the quantum Wakimoto current:
\be \label{qwak}
I = -\left(p^{\bt}(N^{-1}_{\al\bt} g_-^{-1} V^{\al} g_-)\right)
+ g_-^{-1} j_0 g_- +  K g_-^{-1} \dif g_- +
g_-^{-1} \Omega^{\beta} g_- \dif q_{\beta}. 
\ee
Our main result will be  to give the  explicit form of 
the last term, which  represents a quantum correction  due to different
normal orderings.
The function $\Omega^{\beta}(q)$ is  ${\cal G}$-valued and
we inserted some factors of $g_-$ around it for convenience.

To  define  the special  coordinates
in which our formula  for $\Omega^\beta$  will be valid,
we first introduce the matrix  $R^b_a(g_-)$ by 
\be 
g_- T_a g_-^{-1}\equiv R^b_a(g_-)T_b
\qquad g_-\in G_-.
\label{Ad}
\ee 
We  call a  system of global, holomorphic coordinates
$q_\alpha$ on $G_-$  {\em polynomial} if 
$R_a^b(g_-(q))$ is given by a  polynomial  of the  coordinates. 
The fact that   $\det R=1$, which follows from the invariance of $\tr$
and from the fact that $G_-$ is topologically trivial,  
shows that  $R^b_a(g_-^{-1})$  is  also polynomial in the $q_\alpha$.
Furthermore, since one can evaluate $N^{\alpha\beta}(q)$ 
in (\ref{jeq}) using  the adjoint representation,
the definition implies that $N^{\alpha\beta}(q)$ is a polynomial, too.
As it is non-vanishing and is given by a (complex) polynomial,  
the determinant of $N^{\alpha\beta}(q)$ must  be a constant, from
which we see that $N^{-1}_{\alpha\beta}(q)$ is a polynomial as well.
Let now   ${\rm deg}$  denote the gradation with respect to 
which the decomposition (\ref{triang}) was made,  assume that the 
basis elements $V_{\alpha}$ of $\G_-$  
have  well-defined degree, and set 
$d_\alpha \equiv -{\rm deg}(V_{\al}) = {\rm deg}(V^{\al})>0$.
For  polynomial coordinates $q_\alpha$  on $G_-$,  
let us assign degree $d_\alpha$ to $q_\alpha$.
By definition, {\em upper triangular}  coordinates are such 
 polynomial coordinates for which  
$N^{\alpha\beta}(q)$ is given by a homogeneous polynomial of degree 
$(d_\beta-d_\alpha)$ with respect to this assignment of the degree. 
 
In upper triangular coordinates
$N^{\al\bt}(q)$ obviously vanishes  
unless $d_{\bt} \geq d_{\al}$, which  explains  the name
and implies that 
$N^{-1}_{\al\bt}(q)$ is  also a polynomial of  degree $(d_{\bt}-d_{\al})$.
The most obvious examples of upper triangular coordinates are the
`graded exponential coordinates', given by $g_-(q)=\exp(
\sum_{\alpha} q_{\alpha} V_{\alpha} )$. One can also take products
of graded exponential coordinates, by distributing the set $\{ V_{\al} \}$
over disjoint subsets $S_{\cal I}$ and taking
$
g_-(q) = 
\prod_{\cal I} \exp( \sum_{\alpha \in S_{\cal I}} q_{\al} V_{\al} ).
$
If $G=SL(n)$,  
there are even simpler coordinates where the
$q_{\alpha}$ are matrix elements, namely $g_-(q)={\bf 1}_n+\sum_{\alpha}
q_{\alpha} V_{\alpha}$ with $(V_\alpha)_{lk}=\delta_{il}\delta_{jk}$ 
for some $i>j$.
To check the upper triangularity property in these examples,
it is useful to think of $g_-(q)$ as a polynomial
in the $q_\alpha$ and the $V_\alpha$ which are 
declared to have degrees $d_\alpha$ and $-d_\alpha$, respectively.
Then  $g_-(q)$ has  `total degree' zero, and 
$N^{\alpha\beta}(q) V_\beta$ has total degree $-d_\alpha$, 
implying that $N^{\alpha\beta}(q)$ has degree $(d_\beta-d_\alpha)$.

Now we are ready to state our  main result: 

\noindent
{\it Given a system  of upper triangular coordinates $q_{\al}$ on $G_-$, 
the current
${I}$ defined in (\ref{qwak}) satisfies the
OPE given in (\ref{requi}) if 
(i) $2 K_0^0={\vert \psi\vert^2 }(k+h^*)= \vert\psi_i\vert^2 (k_0^i+h_i^*)$ 
for $i>0$,
and (ii) $\Omega^{\beta}$ is given by the following ${\cal G}_-$-valued
object
\be \label{omega}
\Omega^{\beta} = N^{\lambda\rho} \partial^{\beta} N^{-1}_{\gamma\lambda}
 [V^{\gamma},V_{\rho}].
\ee
Furthermore, $\Omega^{\beta}$ is uniquely determined by (\ref{requi})
up to trivial 
redefinitions of the momenta $p^{\beta}$  in (\ref{qwak}),  
\be \label{predef}
p^{\beta} \rightarrow p^{\beta} + (\partial^{\beta} A^{\gamma} - 
\partial^{\gamma} A^{\beta} ) \partial q_{\gamma} 
\ee
with an arbitrary polynomial $A^{\gamma}(q)$. Finally, the stress-energy
tensor for the current ${I}$ takes the form of a free stress-energy
tensor for $p^{\beta},q_{\beta}$ and a sum of improved stress energy tensors
for the currents $j_0^i$ with values in ${\cal G}_0^i$,
\be \label{sugawara}
\frac{1}{2y} \tr ({I}^2)  =
- p^{\beta} \dif q_{\bt} + \frac{1}{2y}\sum_{i\geq 0} \tr(j_0^i j_0^i) 
+\frac{1}{y}\tr((\rho_{{\cal G}} - \rho_{{\cal G}_0} ) \dif j_0)
\ee
where  $y\equiv {1\over 2}\vert \psi\vert^2 (k+h^*)$ and 
$\rho_{{\cal G}} - \rho_{{\cal G}_0} = {1\over 2} [V^{\alpha},V_{\alpha}]$
is half the sum of those positive roots of $\G$ that are 
not roots of $\G_0$.}

The proof of the above statements  consists of an explicit calculation
of the OPEs  of ${I}$ with itself and of the normal
ordered product  $\tr({I}^2)$.
This calculation is  relatively straightforward but  rather long,
and therefore  we do not present it here (see \cite{dBF}).
The relationship between $k$, $k_0^i$ and $K_0^0$ 
given in condition (i) above was also found by Feigin and Frenkel,
although the correct normalization factors are missing in 
Proposition 4 in the second article of \cite{FF}
concerning those cases for which $\vert \psi_i\vert^2 \neq \vert \psi\vert^2$.
We checked that (\ref{qwak}) with (\ref{omega}) reproduces the known
formula  of the currents corresponding to the Chevalley generators of
$\G=sl(n)$ in the principal case.

We wish to note that the  redefinitions of the momenta that
appear in (\ref{predef}) ---which can be absorbed into  redefinitions of 
$\Omega^\beta$ in (\ref{qwak}) leading to  a trivial 
ambiguity  in the solution for $\Omega^\beta$--- 
are  particular quantum 
canonical transformations that preserve the OPEs in (\ref{pqOPE}).
Analogous canonical transformations  of the momenta
at fixed coordinates exist  already at the classical level, 
and actually  no ambiguity arises from the quantization.

More general  canonical transformations of the quantum fields $q_\alpha$,
$p^\beta$ can be used to relate  quantizations of the classical 
Wakimoto realization  in  different systems of coordinates on $G_-$.
Two  systems of coordinates $q_\alpha$ and $Q_\alpha$ on $G_-$
may be called `polynomially equivalent' if the change of coordinates
$q_\alpha=q_\alpha(Q)$ is given by polynomials.
It is shown in \cite{dBF} that 
with any polynomial change of coordinates on $G_-$ one can associate
a canonical transformation of the corresponding set of 
quantum fields, $(q_\alpha, p^\beta) \Leftrightarrow (Q_\alpha, P^\beta)$.
Such a  canonical transformation  naturally induces a  quantization  of the
classical Wakimoto realization in any system of coordinates which
is polynomially equivalent to an upper triangular system of coordinates.
Conversely, the quantizations obtained in polynomially equivalent 
systems of coordinates are always related by quantum canonical 
transformations. 
Finally,  it is also demonstrated in \cite{dBF} that 
the quadratic form of the affine-Sugawara stress-energy tensor
 given in (\ref{sugawara}) 
is maintained under the polynomial changes  of coordinates.  

It is clear that the  polynomiality of  the coordinates 
is necessary for the polynomiality of $I(q,p,j_0)$ in (\ref{qwak}).
We see from the above remarks that  upper triangularity is  a sufficient but 
not a necessary condition  on the `admissible'  polynomial coordinates 
in which the quantization of the classical Wakimoto realization
can be performed. We believe that all such admissible systems
of coordinates are polynomially equivalent, but did not  prove this yet.

\section{Conclusion}
\setcounter{equation}{0}

Wakimoto realizations play an important role in the representation
theory of affine Lie algebras  as the building blocks of resolutions
of irreducible highest weight representations \cite{FF,BF,BMP}. 
Furthermore, they can be used to compute correlation functions in
the WZNW model \cite{FF,BF,BMP,ATY}, and to obtain free field
realizations of ${\cal W}$-algebras \cite{BO,FF2,Fr,Fi}. 
The so-called screening charges play a  crucial role
in all these applications.
Screening charges are operators that commute   with the Wakimoto current,
and  can be used to build intertwiners between different highest
weight representations of the affine Lie algebra. With a compact
explicit expression for the Wakimoto current at our disposal, it is 
possible to explicitly construct the screening charges as well.
Below we summarize the main ideas of the construction, 
referring to  \cite{dBF} for details and for further discussion.

In keeping with the spirit of the preceding sections,  we wish to 
define the classical version of the  screening charges first.
For this we introduce 
a $G_0$-valued field $h_0$ which is related to the current $j_0$ by
\be j_0 = K h_0^{-1} h_0' . \ee
One should be careful not to confuse $h_0$ with the coordinate
$g_0$ on the constrained phase space that appears in Section~3;
$h_0$ is in general a multiple-valued function on the circle.
In terms of $h_0$, we define the charges  $S_{\xi}$ by
\be S_{\xi} \equiv \int_0^{2\pi} d\si \tr(\xi h_0 j h_0^{-1})
\qquad \forall \xi\in \G.  \ee
We then find that $S_{\xi}$ commutes
with the classical Wakimoto current $I$ in (\ref{affwak})
if and only if 
\be \label{clscr}
\xi \in \G_- \cap
[\G_+,\G_+]^{\perp}.
\ee
The $S_{\xi}$ with $\xi$ in (\ref{clscr}) constitute the classical
 screening charges of the Wakimoto realization. 

We wish to  subsequently  quantize the classical screening charges. 
We here restrict ourselves to the principal case, for which  ${\cal G}_0$ is 
the Cartan subalgebra.
In this case, we can bosonize $j_0$ simply by putting
$j_0=-i \sqrt{y} \partial \phi$, where $\phi$ is a $\G_0$-valued scalar 
field with the OPE
\be
\tr (X_0 \underhook{\phi)(z) \tr(Y_0} \phi)(w)  =
-\tr( X_0Y_0) \log(z-w)
\qquad
X_0, Y_0\in \G_0,
\ee
and $y={1\over 2}\vert \psi \vert^2 (k+h^*)$ as before.
In addition, we assume that the basis $V^{\alpha}$ of $\G_+$ has been
chosen to coincide with root eigenspaces, in particular $V^{\al_l}$
is proportional to the step operator $E_{\al_l}$
for every simple root $\alpha_l$.
Noticing  that at the classical level  $h_0= \exp( -{i\over\sqrt{K}}\, \phi)$
and that the space  in (\ref{clscr})  is now spanned by the  $E_{-\alpha_l}$, 
we obtain the following  screening charges at the quantized level:
\be \label{qscr}
S_l  \equiv \oint \frac{dz}{2\pi i}\Bigl(
 p^{\beta} ((N^{-1}_{\al_l \beta} 
:\exp(-\frac{i}{\sqrt{y}}\, \al_l(\phi) ): )\Bigr)
\qquad \quad
l=1,\ldots, {\rm rank}(\G).
\ee
This  formula for the screening charges  
has  been  known  (see e.g.~eq.~(3.12) in the third  article in \cite{BMP}),
although in a less detailed form and, to our knowledge, 
 without a complete proof except for $\G=sl(n)$.
With our explicit formulas 
the  proof that the  $S_l$ in (\ref{qscr})  indeed commute with the 
Wakimoto current  is just a direct  calculation.

One can  use the principal Wakimoto realization to
obtain a free field realization of the ${\cal W}$-algebra 
associated with  the
principal embedding of $sl_2$ in $\G$, sometimes denoted as
${\cal W}[\G]$. This is done by performing a quantum Drinfeld-Sokolov
reduction of $\widehat{\G}_k$ \cite{FF2,dBT}, 
but now using the Wakimoto realization
of ${\widehat\G}_k$ rather than $\widehat{\G}_k$ itself. 
The computation of the BRST cohomology
is very easy, and one finds that the only object that survives is
the $\G_0$-valued scalar $\phi$. In addition, the screening charges 
of ${\widehat \G}_k$ in  
(\ref{qscr}) are cohomologous to
\be \label{wscr}
{\tilde S}_l\equiv  \oint \frac{dz}{2\pi i}
  :\exp(-\frac{i}{\sqrt{y}}\, \al_l(\phi) ): 
\ee
leading to a description of ${\cal W}[\G]$  as the centralizer of the
screening charges of ${\cal W}[\G]$  in (\ref{wscr}) 
acting on the algebra generated by the free  scalar $\phi$. 

Whereas previously several of the above  statements could only be analyzed
by indirect methods, the explicit expressions presented in this letter
allow for straightforward computations, which we expect 
to prove useful in future applications of the Wakimoto realizations.

\bigskip 

\noindent
{\large \bf  Acknowledgements.}
JdB is supported in  part by the National 
Science Foundation grant PHY93-09888.
L.F.~wishes to thank I.~Tsutsui for encouragement.


\end{document}